\documentclass[fleqn,usenatbib,useAMS]{mnras}
\usepackage{mathptmx}
\usepackage[T1]{fontenc}
\usepackage{ae,aecompl}

\usepackage{graphicx}   
\usepackage{amsmath}    
\usepackage{amssymb}    

\newcommand{\kms}{km\,s$^{-1}$}

\title[The magnetic field of Przybylski's star]{
Magnetic and pulsational variability of Przybylski's star (HD\,101065)}

\author[S.~Hubrig et al.]{
S.~Hubrig$^{1}$\thanks{E-mail: shubrig@aip.de},
S.~P.~J\"arvinen$^{1}$,
J.~Madej$^{2}$,
V.~D.~Bychkov$^{3}$,
I.~Ilyin$^{1}$,
M.~Sch\"oller$^{4}$,
\and L.~V.~Bychkova$^{3}$
\\
$^{1}$Leibniz-Institut f\"ur Astrophysik Potsdam (AIP), An der Sternwarte~16, 14482~Potsdam, Germany\\
$^{2}$Astronomical Observatory, University of Warsaw, Al.~Ujazdowskie~4, 00-478~Warszawa, Poland\\
$^{3}$Special Astrophysical Observatory of the Russian Academy of Sciences (SAO), Nizhnij Arkhyz~369167, Russia\\
$^{4}$European Southern Observatory, Karl-Schwarzschild-Str.~2, 85748 Garching, Germany
}

\date{Accepted XXX. Received YYY; in original form ZZZ}

\pubyear{2017}

\begin{document}
\label{firstpage}
\pagerange{\pageref{firstpage}--\pageref{lastpage}}
\maketitle

\begin{abstract}
Since its discovery more than half a century ago Przybylski's star (HD\,101065) continues to excite the
astronomical community by the unusual nature of its spectrum, exhibiting exotic 
element abundances. This star was also the first magnetic chemically peculiar A-type star for which 
the presence of rapid oscillations was established. Our analysis of newly acquired and historic longitudinal 
magnetic field measurements indicates that Przybylski's star is also unusual with respect to its 
extremely slow rotation. 
Adopting a dipolar structure for the magnetic field and using a 
sine wave fit to all reported longitudinal magnetic field values over the last 43\,yr, we find a probable rotation period 
$P_{rot}\approx188$\,yr, which however has to be considered tentative as it does not represent a unique 
solution and has to be verified by future observations.
Additionally, based on our own spectropolarimetric material obtained with HARPS\-pol, we 
discuss the impact
of the anomalous structure of its atmosphere, in particular of the non-uniform horizontal and vertical 
distributions of chemical elements on the magnetic field measurements and the pulsational variability.
Anomalies related to the vertical abundance stratification of Pr and Nd are for the 
first time used to establish the presence of a radial magnetic field gradient.
\end{abstract}


\section{Introduction}
\label{sec:intro}

HD\,101065 was recognized first by \citet{Przybylski1961}
and became one of the most fascinating objects due to its exotic element abundances,
such as a strong Fe underabundance and a large excess of almost all rare earth elements (REE).
HD\,101065 is generally referred to as Przybylski's star.
Using wavelength coincidence statistics, \citet{Cowley2004} suggested the presence of \ion{Pm}{i}, \ion{Pm}{ii},
\ion{Tc}{i}, and possibly \ion{Tc}{ii} in the spectrum of this star.
HD\,101065 is a main-sequence cool
chemically peculiar A type star, a so-called Ap star, and belongs to the group
of rapidly oscillating Ap (roAp) stars,
which pulsate in high radial overtone
$p$ modes with periods in the range 6--24 min.
The presence of rapid oscillation with a period of 12.14\,min was detected by \citet{Kurtz1979}.

A strong longitudinal magnetic field up to $-$2.5\,kG was for the first time measured in 1974 by
\citet{wolff1976}. The average value for the longitudinal field of $-$2.2 kG was obtained using three measurements 
based on 
Zeeman spectra recorded on photographic
plates. The random standard deviation of each field determination was
estimated as 450\,G. \citet{wolff1976} noted that there is no 
evidence for a variability of the magnetic field strength over a time interval of one year.

The magnetic fields in Ap stars
can be described with the oblique rotator model \citep{stibbs1950}.
To a first approximation their fields can be modeled with a simple
dipole structure with the axis of the magnetic dipole inclined to the
rotation axis. To determine the rotation period in Ap stars, apart from photometric 
variability studies, spectropolarimetric and spectroscopic  studies are widely used to monitor the
variation of the longitudinal magnetic field or the magnetic field modulus, measured in 
magnetically resolved lines. 
According to \citet{Mathys2015}, Ap stars have rotation periods that span 5 to 6 
orders of magnitude, from about 0.5\,d to about 300 years or even longer, and as of today, the 
long-period tail of the distribution of the rotation periods of Ap stars
is populated almost exclusively by stars that show magnetically
resolved lines. Magnetically split components are usually observed in stars with a
magnetic field modulus down to 2.2\,kG \citep{Mathys1997}.
Since the spectrum of HD\,101065 exhibits rather narrow spectral
lines with $v\,\sin\,i = 3.5\pm0.5$\,km\,s$^{-1}$ \citep{Cowley2000}, it was possible to measure a weak 
magnetic broadening corresponding to 2.3\,kG obtained from partially resolved \ion{Gd}{ii} and 
\ion{Sm}{ii} lines \citep{Cowley2000}. According to  \citet{Mkrtichian2008}, HD\,101065 appears to be a prime
candidate to possess a very long rotation period. 

To study the periodicity of
the magnetic variability of this star we carried out over the last three years high-resolution 
spectropolarimetric observations using the High Accuracy Radial velocity Planet Searcher polarimeter
\citep[HARPS\-pol;][]{snik2008} 
attached to ESO's 3.6\,m telescope (La Silla, Chile).
We complemented the measured longitudinal magnetic field strengths with
magnetic measurements from the literature to characterize the field variation curve.
Additionally, since magnetic Ap stars and, in particular, roAp stars frequently display non-uniform horizontal and 
vertical distributions
of chemical elements, we investigate the possible impact of such atmospheric properties on the measured magnetic 
field strength and the pulsational characteristics.

\section{Observations and data reduction}
\label{sec:obs}

\begin{table}
\centering
\caption{
Logbook of the HARPSpol observations. 
}
\label{tab:log}
\begin{tabular}{ccc}
\hline
\multicolumn{1}{c}{HJD} &
\multicolumn{1}{c}{Date} &
\multicolumn{1}{c}{$S/N$} \\
\hline
2457179.4993 & 2015--06--05 & 314  \\
2457555.6030 & 2016--06--16 & 377 \\
2457908.5970 & 2017--06--04 & 232 \\
2457911.6047 & 2017--06--07 & 351 \\
\hline
\end{tabular}
\end{table}

HARPSpol observations using the circular polarization analyzer were obtained on the nights of
2015 June 5, 2016 June 16, and 2017 June 4 and 7.
The recorded spectra cover the 3780--6910\,\AA{} wavelength range with a small gap between 
5259 and 5337\,\AA{} at a 
spectral resolution of $R\approx115\,000$.
We observed the star with a sequence of two sub-exposures in the first three occasions
and of four sub-exposures during the last epoch, obtained rotating the quarter-wave retarder 
plate by 90$^\circ$ after each sub-exposure.  
The reduction and calibration of these spectra was performed using the HARPS data
reduction software available on La~Silla. The normalization of the spectra to
the continuum level was described in detail by 
\citet{Hubrig2013}. The dates of observations and the achieved 
signal-to-noise ratios in the Stokes~$I$ spectra are presented in Table~\ref{tab:log}.

\section{Magnetic field analysis}
\label{sec:mfield}

\subsection{Measurements using the least-squares deconvolution technique} 
\label{sec:lsd}

To measure the longitudinal magnetic field, we employed 
the least-squares deconvolution (LSD) technique, allowing
us to achieve a much higher signal-to-noise ratio ($S/N$) in the LSD spectra.  LSD combines line 
profiles (assumed to be identical) centred on the position of the 
individual lines and scaled according to the line strength and 
sensitivity to a magnetic field (i.e.\ line wavelength and Land{\'e} 
factor). The details of
this technique and of the calculation of the Stokes~$I$ and $V$
parameters can be found in the work of \citet{Donati1997}. 

Magnetic Ap stars frequently display non-uniform horizontal and vertical distributions 
of chemical elements.
Some elements show preferential surface concentration close to the magnetic poles, while other elements
are concentrated closer to the magnetic equator regions. 
Thus, the lines of different elements with 
different abundance distributions across the stellar surface sample
the magnetic field in a different manner. Combining lines belonging to different elements altogether,
as is frequently done with the LSD technique, may lead to dilution of  the
magnetic signal or even to its (partial) cancellation, if enhancements of 
different elements occur in regions of opposite magnetic polarity.

To investigate the possible impact of such a horizontal non-uniformity, we decided to 
measure the magnetic field using eight different line masks:
one mask Fe, which includes the least blended iron lines,
five individual masks for the best identified lines belonging to lanthanides, such as 
La, Ce, Pr, Nd, and Sm, 
one mask REE that combines all individual masks for the lanthanides, and one mask All that combines the  masks Fe and REE.
These eight line masks were constructed using the Vienna Atomic Line Database
\citep[VALD3;][]{kupka2011} and the stellar
parameters $T_{\rm eff}=6400$\,K and $\log\,g=4.2$ reported for HD\,101065 in the study by \citet{Shulyak2010}.
The LSD Stokes~$I$ and $V$ spectra, calculated for the eight line masks,
were used for the measurement of the longitudinal magnetic
field through the first order moment of the Stokes~$V$ profile 
\citep[e.g.][]{Mathys1989}.

\begin{table}
\centering
\caption{
Longitudinal magnetic field strengths measured at four different observing epochs 
using the LSD technique for the eight different line masks. In Columns~2 and 3 we present the number of lines in the individual
line masks and the average Land{\'e} factor. 
For each line list, the four rows in the last column correspond to the measurements obtained on the observing dates 
2015 June~5,
2016 June~16,
2017 June~4, and
2017 June~7, respectively. In all cases the false alarm 
probability is less than $10^{-15}$.
}
\label{tab:mf}
\begin{tabular}{lrcl@{$\pm$}r}
\hline
\multicolumn{1}{c}{Line mask} &
\multicolumn{1}{c}{\# of} &
\multicolumn{1}{c}{Land\`e} &
\multicolumn{2}{c}{$\left<B_{\rm z}\right>$} \\
\multicolumn{1}{c}{} &
\multicolumn{1}{c}{lines} &
\multicolumn{1}{c}{Factor} &
\multicolumn{2}{c}{(G)} \\
\hline
All       & 417 & 1.160  & $-$732 & 20 \\
          &      &           & $-$734 & 19 \\
          &      &           & $-$738 & 20 \\
          &      &           & $-$735 & 19 \\
\hline
Fe        & 23 & 1.211   & $-$722 & 62 \\
          &      &           & $-$737 & 63 \\
          &      &           & $-$708 & 62 \\
          &      &           & $-$720 & 61 \\
\hline
REE       &  394 & 1.157 & $-$751 & 19 \\
          &      &           & $-$752 & 19 \\
          &      &           & $-$757 & 19 \\
          &      &           & $-$754 & 19 \\
\hline
La        &   47 & 1.072 & $-$652 & 52 \\
          &      &           & $-$658 & 51 \\
          &      &           & $-$668 & 53 \\
          &      &           & $-$660 & 51 \\
\hline
Ce        &   57 & 1.059 & $-$609 & 78 \\
          &      &           & $-$616 & 80 \\
          &      &           & $-$626 & 75 \\
          &      &           & $-$620 & 76 \\
\hline 
Pr        &   84 & 1.116 & $-$636 & 68 \\
          &      &           & $-$636 & 71 \\
          &      &           & $-$668 & 71 \\
          &      &           & $-$642 & 72 \\
\hline 
Nd        &  145 & 1.134 & $-$856 & 32 \\
          &      &           & $-$863 & 31 \\
          &      &           & $-$865 & 32 \\
          &      &           & $-$861 & 30 \\
\hline
Sm        &   61 & 1.424 & $-$713 & 31 \\
          &      &           & $-$709 & 32 \\
          &      &           & $-$717 & 31 \\
          &      &           & $-$714 & 33 \\
\hline
\end{tabular}
\end{table}

In Table~\ref{tab:mf} we present the number of lines in each line mask, 
the average Land{\'e} factors and the measured longitudinal magnetic field values
for the observations obtained between 2015 and 2017. In all cases the false alarm 
probability (FAP) is less than $10^{-15}$. According to the convention of \citet{Donati1992},
a Zeeman profile with FAP\,$\le$\,$10^{-5}$ is considered 
as a definite detection, $10^{-5}$\,$<$\,FAP\,$<10^{-3}$ as a marginal detection, and 
FAP\,$>$\,$10^{-3}$ as a non-detection.

\begin{figure*}
\centering
\includegraphics[width=1.00\textwidth]{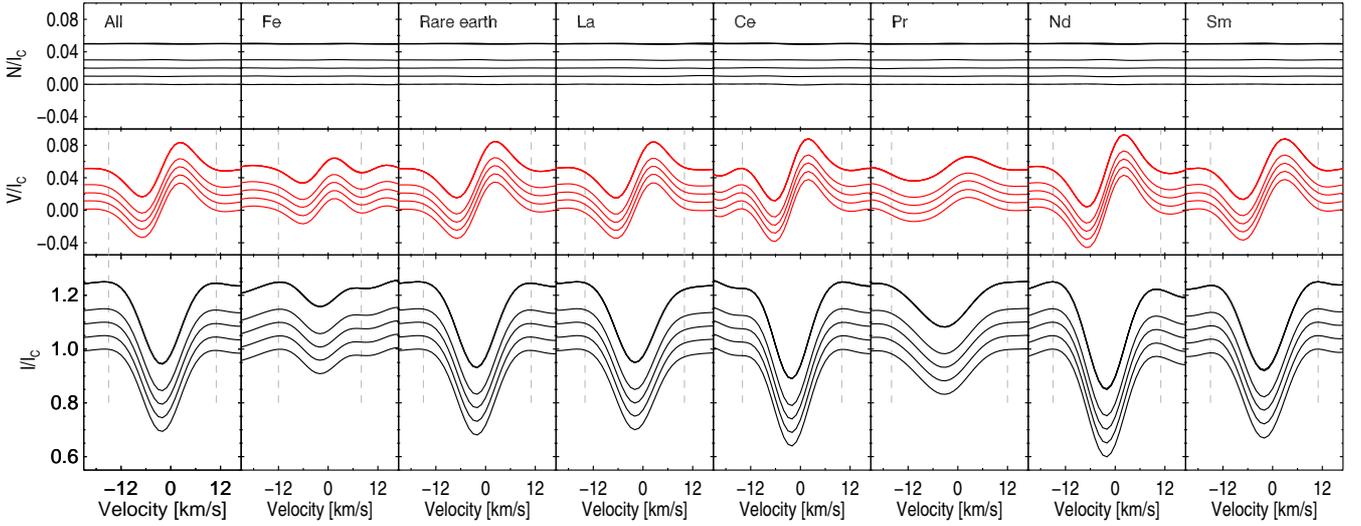}
\caption{
  LSD Stokes~$I$ (bottom), Stokes~$V$ (middle), and diagnostic null ($N$) spectra
  (top) obtained for HD\,101065 on four nights between 2015 and 2017.
The LSD spectra were calculated using eight different
  line lists, indicated in each panel.
For each panel, spectra for 2015 June 5, 2016 June 16, 2017 June 4, and 2017 June 7 
are shown from bottom to top; the last three are shifted upwards for better visibility.
The LSD Stokes~$I$, Stokes~$V$, and diagnostic null ($N$) spectra
  at the top show all four spectra overplotted.
}
\label{fig:rot}
\end{figure*}

The LSD Stokes~$I$, $V$, and diagnostic null ($N$) spectra for all line masks are presented 
in Fig.~\ref{fig:rot}. The diagnostic $N$ profiles
are usually used to identify spurious polarization signatures. They are calculated by 
combining the subexposures in such a way that the polarization cancels out.
The essentially flat null spectra presented in Fig.~\ref{fig:rot} indicate the absence of spurious contributions 
to our measurements.

\subsection{Horizontal inhomogeneous element distribution} 
\label{sec:hor}

The LSD Stokes~$I$ spectra in the lower panels of Fig.~\ref{fig:rot} clearly
demonstrate  abundance  excesses  of  REE and a deficiency of iron, already mentioned
in previous works (e.g.\ \citealt{Cowley2000}). Among the REE, the neodymium and cerium lines are 
the strongest
in the spectrum of HD\,101065, whereas the praseodymium  lines appear rather weak.
The LSD Stokes~$V$ spectra show the presence of clear Zeeman features corresponding to a 
longitudinal magnetic field of negative polarity. The field strengths for each 
line mask measured on the four epochs over the three years appear to be constant within  uncertainties.
On the other hand, the measurements using 
individual masks with lines corresponding to different elements reveal distinct differences in
the field strengths, with the lowest field value obtained for cerium,
$\left<B_{\rm z}\right> \approx -620$\,G,
and the strongest field value,  $\left<B_{\rm z}\right> \approx -860$\,G, for neodymium. 
This suggests that the neodymium lines may form in spots that
are close to the magnetic pole, while cerium lines form at some distance from the magnetic pole.
 
Significant differences
in the magnetic field strengths measured using spectral lines belonging to different elements were already 
mentioned in the past
by \citet{Hubrig2004b} who used spectropolarimetric observations of several Ap stars  with the FOcal Reducer 
low dispersion Spectrograph (FORS\,1; \citealt{Appenzeller1998}) mounted on the 8\,m Antu telescope of 
the VLT. The authors reported that the magnetic field strength measured
in Balmer lines can be by up to 500\,G lower compared with measurements where metal lines are included.
The detected differences between measurements carried out using lines of different elements
can most likely be ascribed to significant inhomogeneous distributions of these elements in 
both horizontal and vertical directions, indicating a very peculiar atmospheric structure
in this star.

\subsection{Vertical stratification of Pr and Nd}
\label{sec:vert}

Most of the studied roAp stars show a REE anomaly, i.e.  abundances derived from REE lines 
of different
ionization  stages show differences of up to a few dex (e.g.\ \citealt{Cowley2000,Ryabchikova2004}).   
It is generally assumed that a strong magnetic field suppresses convection and provides
a stable environment where radiative levitation of some elements occurs. This radiative diffusion
mechanism leads to a stratified atmosphere where the abundance distribution in vertical direction can 
for simplicity be approximated
by a step function with a different element abundance in the higher atmospheric layer compared to that in the 
deeper layer (e.g.\ \citealt{Shulyak2009}).

The REE anomaly in HD\,101065 was for the first time discovered by \citet{Cowley2000} and it was suggested
that due to the influence of radiative diffusion, REE are accumulated in the upper atmospheric layers.
According to \citet{Shulyak2010}, due to the heavy blending of REE lines in the spectrum of this star,
their stratification analysis is highly complicated. The authors emphasize that similar to the hotter roAp star 
HD\,24712 with a strong accumulation of REE in the upper atmosphere  producing a characteristic inverse 
temperature gradient, one would expect the same mechanism to operate in the atmosphere
of HD\,101065. In an attempt to construct an empirical self-consistent model atmosphere for HD\,24712, 
where the stratification of chemical elements is derived directly from the observed
spectra and then treated in a model atmosphere code, \citet{Shulyak2009} report that
the strong overabundance of Pr and Nd leads to the appearance of an inverse temperature gradient with a maximum
temperature increase of up to 600--800\,K compared to a homogeneous abundance model.
The spectra of many roAp stars also show a strong core-wing anomaly in the hydrogen lines, particularly the H$\alpha$
line (e.g.\ \citealt{Cowley2001}). This anomaly is usually explained by non-standard temperature gradients.

\begin{figure}
\centering
\includegraphics[width=.48\textwidth]{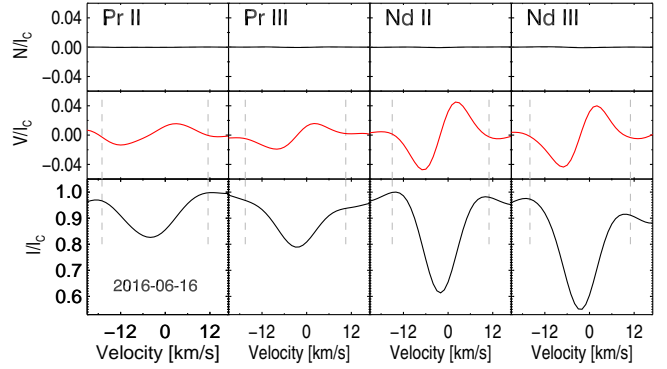}
\caption{
  As Fig.~\ref{fig:rot} but using the \ion{Pr}{ii}, \ion{Pr}{iii}, \ion{Nd}{ii},
  and \ion{Nd}{iii} lines in the spectra obtained on 2016 June~16.
}
\label{fig:PrNd23IVN}
\end{figure}

The anomalous structure of the atmosphere of HD\,101065 is also evident in our observations. As an example, we
present in Fig.~\ref{fig:PrNd23IVN}
the LSD Stokes~$I$, $V$, and diagnostic null ($N$) spectra for observations in 2016 obtained with the highest
$S/N$ using four masks containing 
lines belonging to the first and second ionization stages of Pr and  Nd.
These elements have a larger number of 
identified lines compared to other REE and thus are best suited for our analysis. 
The visual inspection of the LSD Stokes~$I$ profiles shows that Pr and  Nd lines in 
the second ionization stage appear stronger than Pr and Nd lines in the first ionization stage. 
According to \citet{Mashonkina2005,Mashonkina2009}, the line formation of Pr and Nd can strongly
deviate from the local thermodynamic equilibrium (LTE) and the doubly ionized lines of these elements are unusually strong
due to combined effects of vertical stratification and departures from LTE.

It is noteworthy that REE anomalies related to vertical abundance stratification can be used 
to establish
the presence of the magnetic field gradient in the stellar atmospheres. Using the line mask 
for the \ion{Nd}{ii} lines we measure in the spectra obtained on 2016 June 16 a longitudinal magnetic 
field $\left<B_{\rm z}\right>=-970\pm38$\,G
whereas using the line mask for the \ion{Nd}{iii} lines we obtain $\left<B_{\rm z}\right>=-667\pm32$\,G. The effect of the
magnetic field gradient is less noticeable for Pr, where
we measure for the \ion{Pr}{ii} lines $\left<B_{\rm z}\right>=-689\pm102$\,G, and $\left<B_{\rm z}\right>=-588\pm80$\,G
using the line mask for the \ion{Pr}{iii} lines.
Identical results on the presence and strength of a magnetic field gradient
were also obtained using the lines belonging to the REE in the individual observations acquired
in 2015 and 2017.
To increase the accuracy of our measurements, we also calculated mean HARPSpol spectra representing the average 
polarimetric spectra over all three years and remeasured
the magnetic field using the Pr and Nd lines of different ionization stages. We obtain $\left<B_{\rm z}\right>=-969\pm17$\,G 
using the \ion{Nd}{ii} lines and $\left<B_{\rm z}\right>=-663\pm18$\,G using the \ion{Nd}{iii} lines, whereas using the line mask for
the \ion{Pr}{ii} lines we measure $\left<B_{\rm z}\right>=-690\pm35$\,G and $\left<B_{\rm z}\right>=-596\pm28$\,G using
the \ion{Pr}{iii} lines. 

The large difference in the magnetic field strengths measured on the Nd lines of different ionization stages
with the stronger magnetic field detected in the \ion{Nd}{ii} lines and a weaker magnetic field in the \ion{Nd}{iii} lines is highly 
intriguing and unexpected.
For a non-peculiar A-type star, we would expect
lines of Nd and Pr of the second ionization stage to be formed lower in the atmosphere than
lines of Nd and Pr of the first ionization stage.
Assuming a dipole configuration of the magnetic field,
we would then expect to measure a stronger magnetic field in the
lines of Nd~{\sc iii} and Pr~{\sc iii}.
However, assuming a normal atmospheric structure, our measurements show a significant
decrease of the magnetic field strength with atmospheric depth.
The measured inverse magnetic field gradient can
most probably be explained by non-standard temperature gradients in the atmospheres of roAp stars.

In line with these results, \citet{Nesvacil2004} found a significant decrease of the  
magnetic field strength,  of the order of a few hundred  Gauss,  with atmospheric depth in a few Ap stars
using measurements of the mean magnetic field modulus from spectral lines resolved into
magnetically split components lying on different sides of the Balmer jump.
One star in their sample, the roAp star 33\,Lib, showed the largest difference (up to 6$\sigma$)
in the mean magnetic field modulus measured at different atmospheric depths.
However, estimations of the optical depth for the analyzed spectral
lines in that work were carried out using atmosphere models for normal non-magnetic stars and certainly do not
correspond to realistic optical depths, which are expected to be different in roAp stars due to the 
appearance of an inverse temperature gradient. Indeed, \citet{Kurtz2004} studied the atmospheric depth dependence
of pulsations in 33\,Lib as a function of atmospheric depth and suggested that \ion{Nd}{ii} and \ion{Nd}{iii} lines
form on opposite sides of a pulsation node with \ion{Nd}{iii} above
and \ion{Nd}{ii} below the node. Also the work of \citet{Mkrtichian2003} on radial velocity variations in 
these stars indicates that \ion{Nd}{iii} lines are formed significantly higher in the stellar atmosphere 
than \ion{Nd}{ii}. Since HD\,101065 is a typical roAp star, we expect  neodymium  stratification  with 
\ion{Nd}{iii} lines formed in the upper layer of this star. Thus, the stronger magnetic field
measured using the \ion{Nd}{ii} lines corresponds to a larger atmospheric depth as expected for a dipole
configuration of the magnetic field.

In view of the extremely abnormal chemical composition of the atmospheres and the presence of strong magnetic fields,
the calculation of the element abundance distribution with respect to the optical depths in roAps is a very complex
process requiring the fitting of spectroscopic, photometric, and magnetic data. The vertical stratification  of  
the chemical elements was modeled in previous studies of roAp stars in a simplified way, assuming a
simple step-like profile of the distribution of elements. However, in realistic self-consistent 
atmospheric models the stratification and abundance analysis should be
linked to the model atmosphere calculation. Obviously,
significant improvements in the physics of atmospheric 
models and in opacity sources, also taking into account realistic magnetic-field configurations, are urgently needed.

\subsection{Application of the moment technique}
\label{sec:moment}

We also analysed the HARPS spectropolarimetric material 
with a different approach for the
measurements of the magnetic field, namely the moment technique
developed by \citet{Mathys1991,Mathys1995}. 
This technique allows us not only to determine the mean longitudinal magnetic
field, but also to prove the presence of the crossover effect and measure the
quadratic magnetic field. This information cannot be obtained
from the LSD technique. 

The mean quadratic magnetic field is derived through the application of the moment technique, 
described e.g.\ by \citet{Mathys2006}. It is determined from the study of the 
second-order moments of the line profiles recorded in unpolarized 
light (that is, in the Stokes parameter $I$): 

\begin{equation}
\langle B_q\rangle= (\langle B^2\rangle + \langle B_z^2\rangle)^{1/2},
\end{equation}

\noindent
where $\langle B^2\rangle$ is 
the mean square magnetic field modulus, i.e.\ the average over the stellar 
disc of the square of the modulus of the magnetic field vector, weighted by 
the local emergent line intensity, while $\langle B_z^2\rangle$ is 
the mean square longitudinal magnetic field, i.e.\ the average over the stellar 
disc of the square of the line-of-sight component of the magnetic 
vector, weighted by the local emergent line intensity. 
The crossover effect can be measured by the second order moment about their centre of the profiles of 
spectral lines recorded in the Stokes parameter $V$.  
It was for the first time shown by \citet{Mathys1995} that it is possible to derive from the measurements a quantity called the 
mean asymmetry of the longitudinal magnetic field, which is the first moment of the component of the 
magnetic field along the line of sight, about the plane defined by the line of sight and the stellar 
rotation axis.

We employed in the analysis the same 23 least blended \ion{Fe}{i} lines as those used in the LSD method.
In agreement with the LSD results, our measurements of HARPSpol spectra reveal the longitudinal
magnetic field  $\left<B_{\rm z}\right>=-757\pm63$\,G at a significance level of 12$\sigma$.
We do not detect any significant crossover on the four observing epochs, as expected
for stars with very long rotation periods, but we detect a mean quadratic
magnetic field of the order of 1.6$\pm$0.3\,kG. 
No significant longitudinal fields were determined from the diagnostic null spectra.

\section{Magnetic field variability}
\label{sec:per}

\begin{table}
\centering
\caption{
Historic magnetic field measurements.
In the first column we list the heliocentric Julian date at the time of observation, followed by the
longitudinal magnetic field value. The references to the related studies are presented
in the last column.}
\label{tab:mf_prev}
\begin{tabular}{ll@{$\pm$}rc}
\hline
\multicolumn{1}{c}{HJD} &
\multicolumn{2}{c}{$\left<B_{\rm z}\right>$} &
\multicolumn{1}{c}{Reference}\\
\multicolumn{1}{c}{} &
\multicolumn{2}{c}{(G)} &
\multicolumn{1}{c}{} \\
\hline
2442386.22 & $-$1567 & 210&\citet{wolff1976}\\
\hline
2448782.54 & $-$1408 &  50&\citet{Cowley1998} \\
\hline
2452383.20 & $-$1041 &  53&\citet{Hubrig2004b}\\
2452701.25 & $-$1004 &  75& \\
\hline
2454209.76 & $-$1107 &  19&Hubrig, unpublished \\
2454222.60 & $-$1071 &  19& \\
2454233.49 & $-$1046 &  20& \\
2454247.47 & $-$1086 &  20& \\
2454254.57 & $-$1051 &  20& \\
2454272.53 & $-$1036 &  19& \\
2454281.48 & $-$1012 &  20& \\
2454297.52 & $-$1024 &  20& \\
2454306.54 &  $-$932 &  26& \\
\hline
\end{tabular}
\end{table}

The fact that the longitudinal magnetic field strengths 
for each line mask measured on four epochs over the last three years are constant within the measurement uncertainties 
indicates that the rotation period of HD 101065 is long.
In Table~\ref{tab:mf_prev} we list all available longitudinal magnetic field measurements published in the 
literature. 
We also present a number of FORS\,1 longitudinal magnetic field measurements that we did not publish earlier.
The dispersion of the longitudinal magnetic field measurements obtained with FORS\,1
in spectropolarimetric mode attached to one 
8\,m telescope of the VLT can likely be explained by the rather short exposure time of each observation consisting of a
sequence of eight subexposures taken at two different position angles of the retarder waveplate,
$+45^{\circ}$  and $-45^{\circ}$ \citep[e.g.][]{Hubrig2004a}, and of a duration of only  65\,s.
The duration of the full sequence of about 10\,min is less than 
the length of the pulsation period of HD\,101065 of 12.4\,min. Hence an impact of pulsational variability on our FORS\,1 
measurements cannot be excluded. From the theoretical considerations presented by \citet{Hubrig2004b} follows that 
a pulsationally-modulated variation of the order of 100\,G may exist in the outer atmospheric layers of roAp stars with kG
magnetic fields.
We note that due to a lack of good atomic data used for the calculation of the Land\'e factors 
in the study of \citet{wolff1976}, we adopted the revised longitudinal magnetic field value 
$\left<B_{\rm z}\right>=-1567\pm210$\,G reported by \citet{Mkrtichian2008}.
These measurements including our most recent HARPSpol measurements clearly 
indicate a gradual decrease of the longitudinal magnetic field strength \citep[see also e.g.][]{Mkrtichian2008}.

\begin{figure}
\centering
\includegraphics[width=0.48\textwidth]{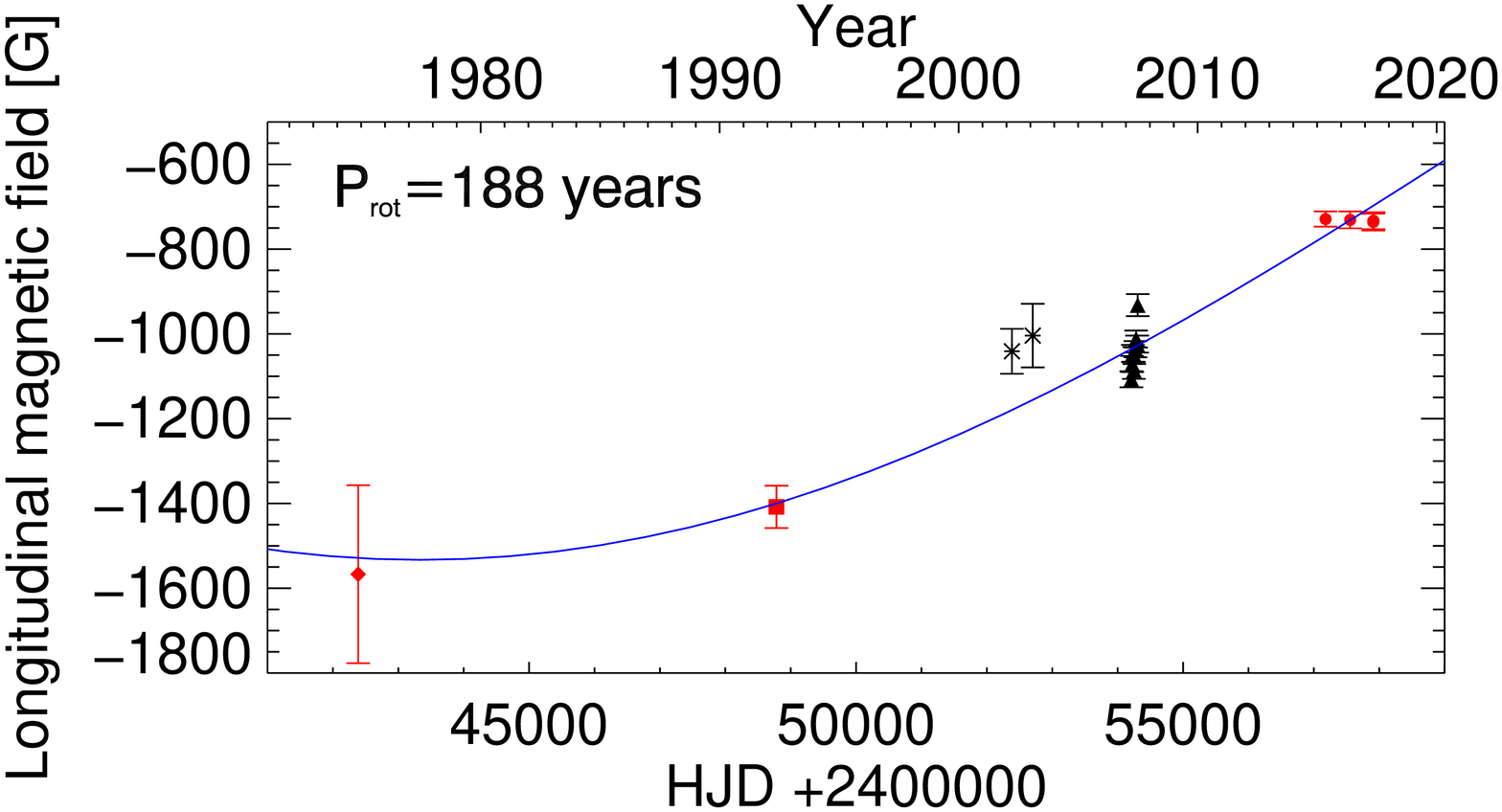}
\caption{
Variation of the longitudinal magnetic field values $\left<B_{\rm z}\right>$
for HD\,101065 as a function of 
HJD between 1974 and 2017. The photographic measurements by \citet{wolff1976} and the 
high-resolution CCD measurements are highlighted in red (in the online version), whereas
the low-resolution FORS\,1/2 measurements are indicated in black color. Vertical error bars show
the measurement accuracies. 
}
\label{fig:per}
\end{figure}

\begin{figure}
\centering
\includegraphics[width=0.48\textwidth]{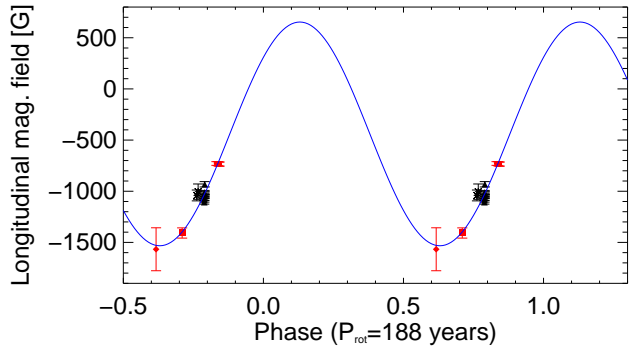}
\caption{
The longitudinal magnetic field values $\left<B_{\rm z}\right>$
for HD\,101065 measured over 43\,yr and phased with the period of 188\,yr.
}
\label{fig:trend}
\end{figure}

In Fig.~\ref{fig:per} we plot the longitudinal magnetic field values 
for HD\,101065 as a function of HJD between 1974 and 2017.
Adopting a dipolar structure for the magnetic field and using a sine wave fit to all reported field values, 
a minimum $\chi^2$ solution of 4.6 utilizing the Levenberg--Marquardt method \citep{Press1992}
was found for a probable rotation period of $P_{rot}\approx188$\,yr. 
In Fig.~\ref{fig:trend} we present 
the observed trend in the magnetic field measurements over 43\,yr together with the full sine wave
computed for this period.
Obviously, since only 23\% of the period of about 188\,yr is covered by the magnetic field measurements,
this very long period should be 
considered tentative and has to be verified by additional observations in the next tens of years.

\section{Pulsational variability}
\label{sec:oscil}

The roAp stars were the first stars for which high-overtone p-mode pulsations were detected.
Previous pulsational studies
using high-resolution spectroscopic time series also showed that
pulsation radial velocity (RV) amplitudes are different for different elements, with the highest
amplitudes detected in REE
\citep[e.g.][]{Elkin2010,Elkin2015}.
HD\,101065 is reported to show one of the highest photometric pulsation
amplitudes
\citep{Kurtz2006}
and rather high pulsation RV amplitudes of almost 0.9\,\kms{}
for lines of \ion{Pr}{iii} and up to about 0.2\,\kms{} for \ion{Ce}{ii}
lines
\citep{Elkin2015}.

\begin{figure*}
\centering
\includegraphics[width=0.49\textwidth]{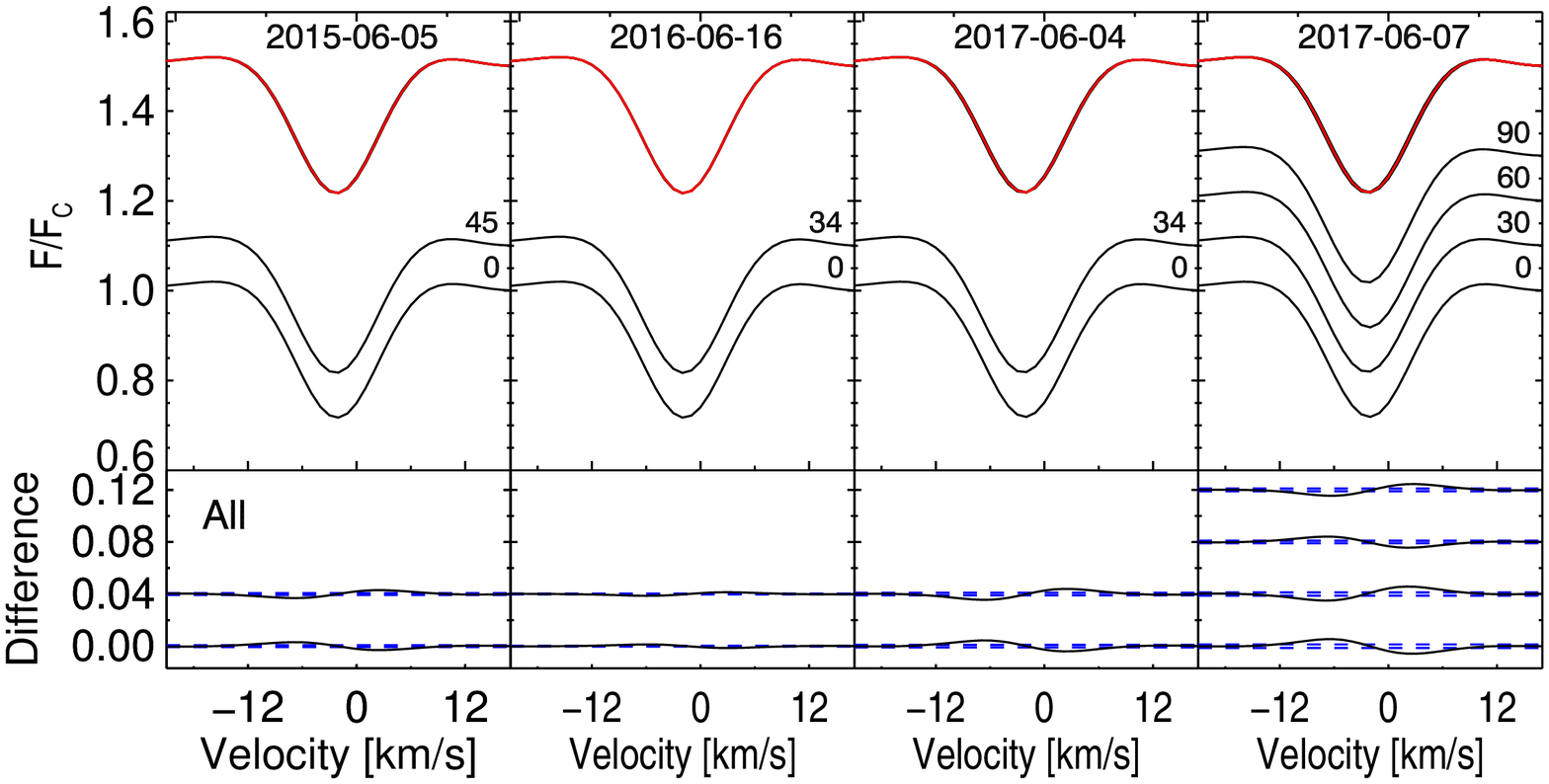}
\includegraphics[width=0.49\textwidth]{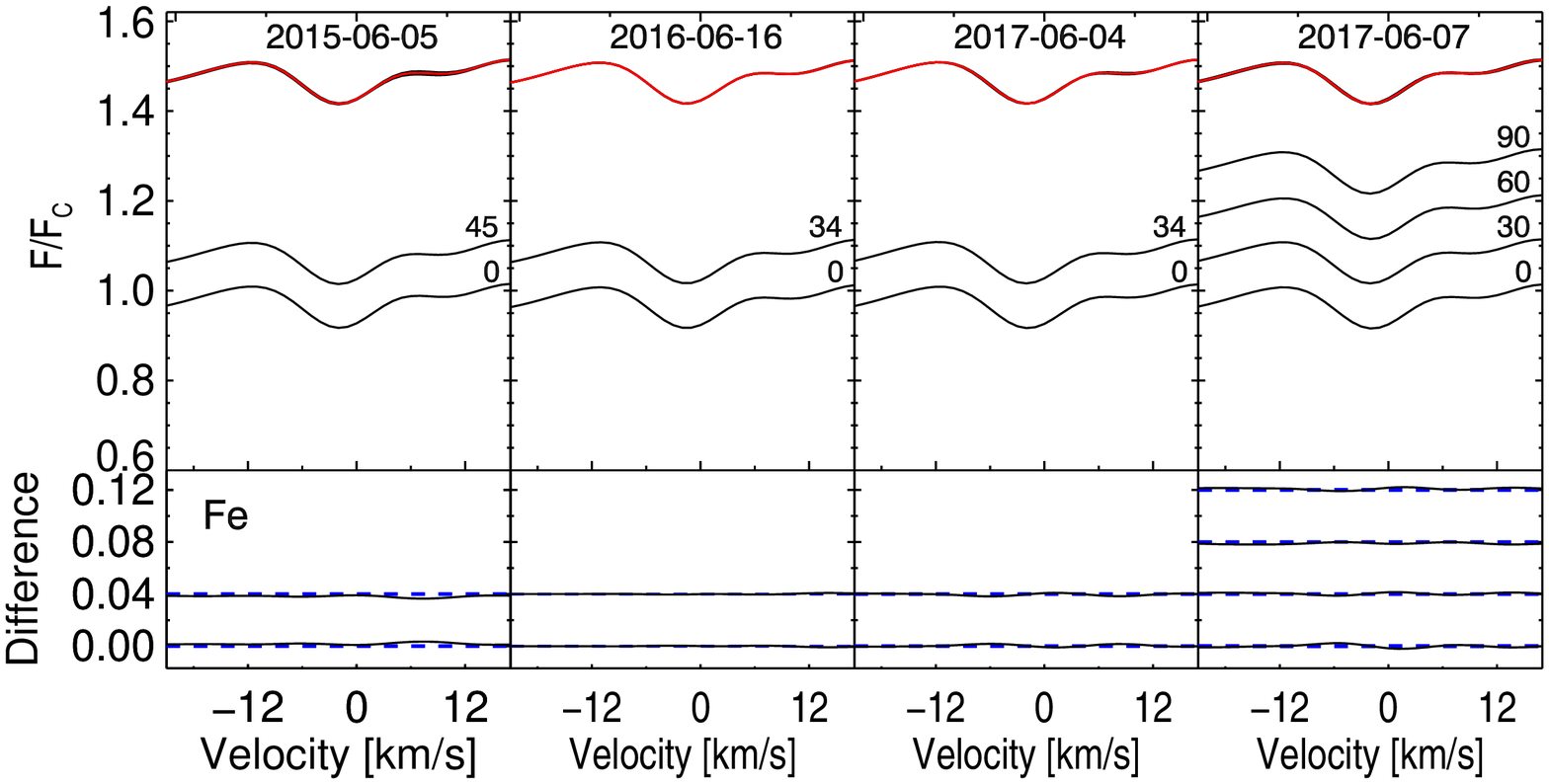}
\includegraphics[width=0.49\textwidth]{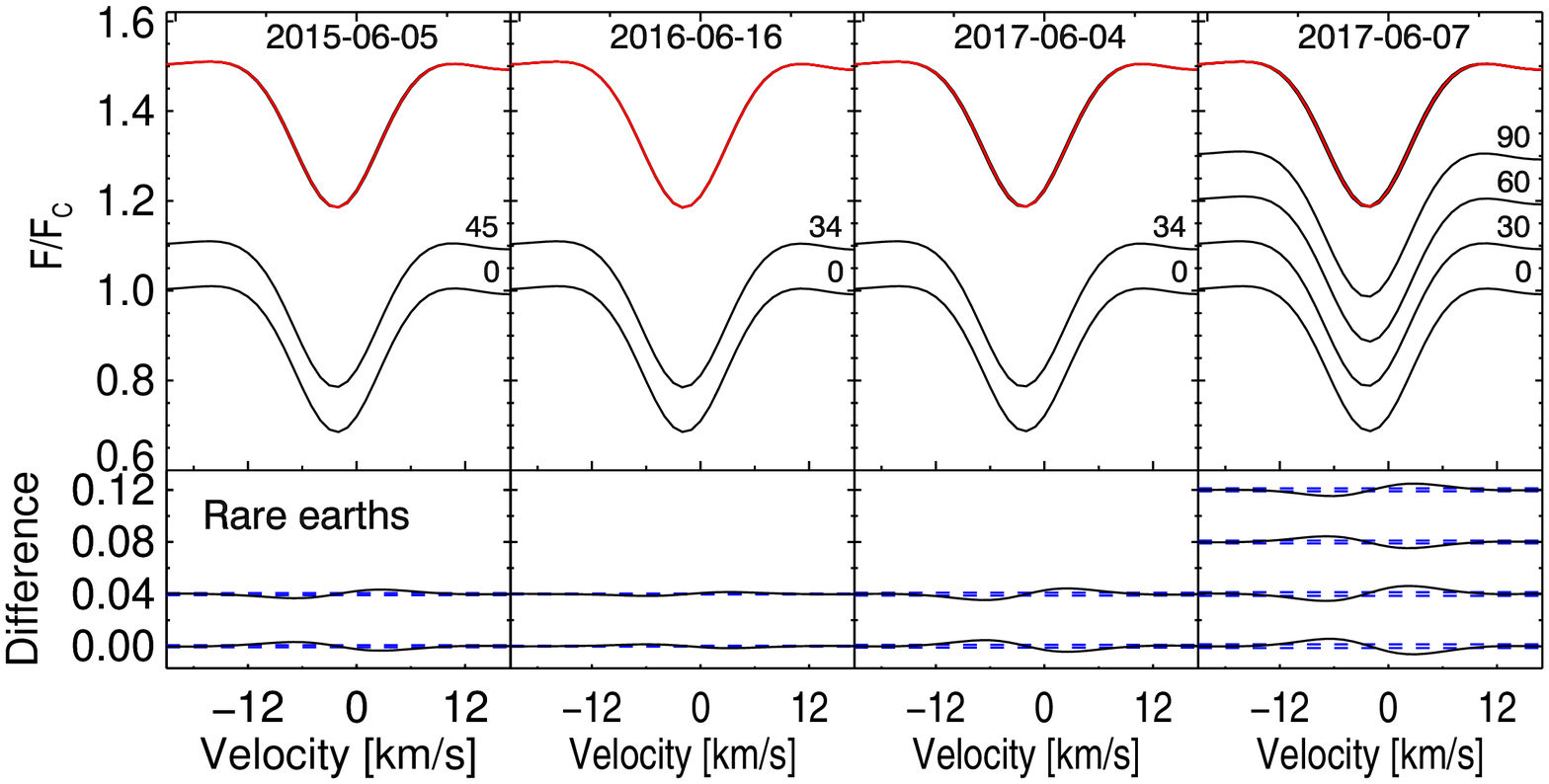}
\includegraphics[width=0.49\textwidth]{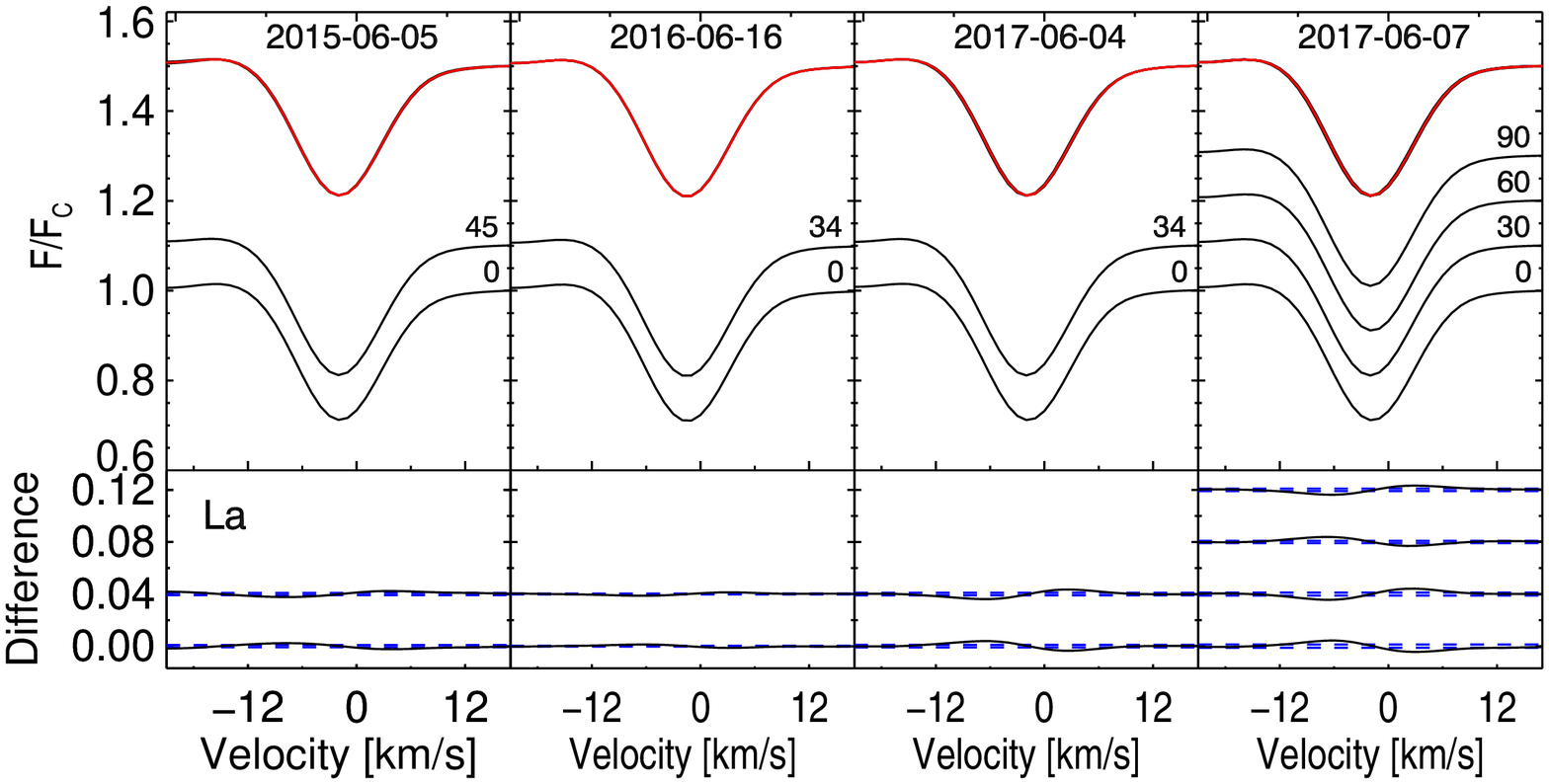}
\includegraphics[width=0.49\textwidth]{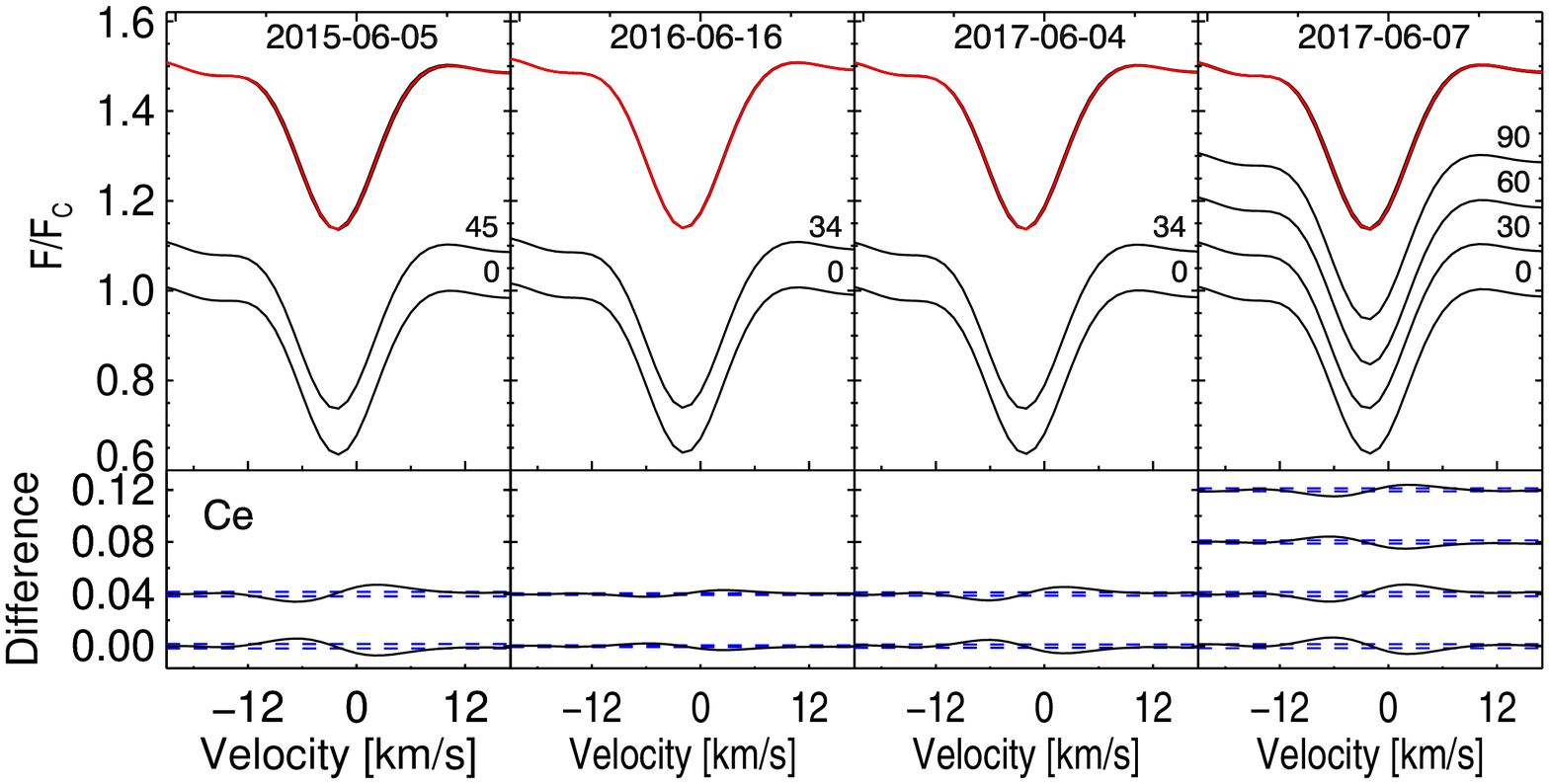}
\includegraphics[width=0.49\textwidth]{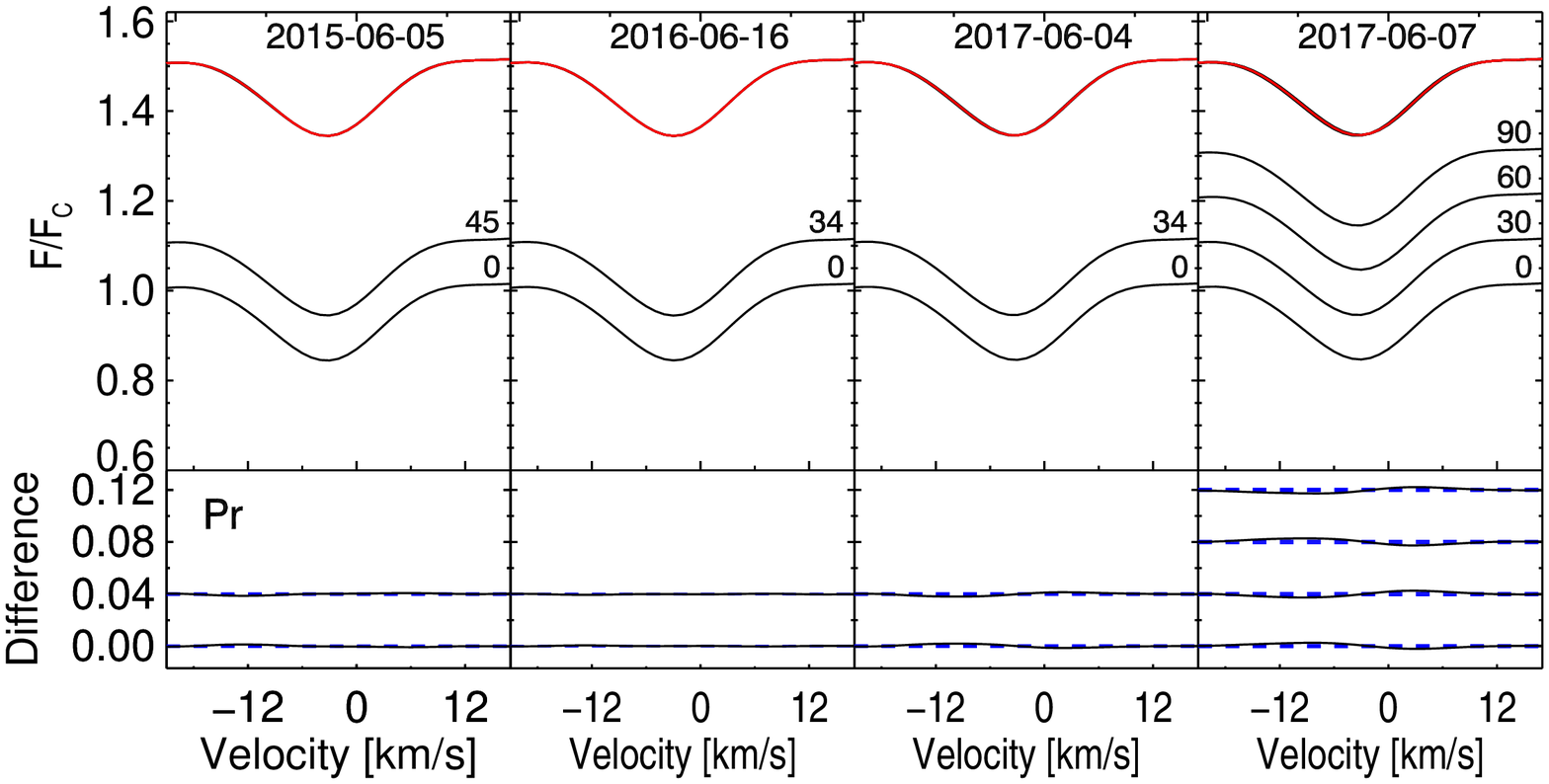}
\includegraphics[width=0.49\textwidth]{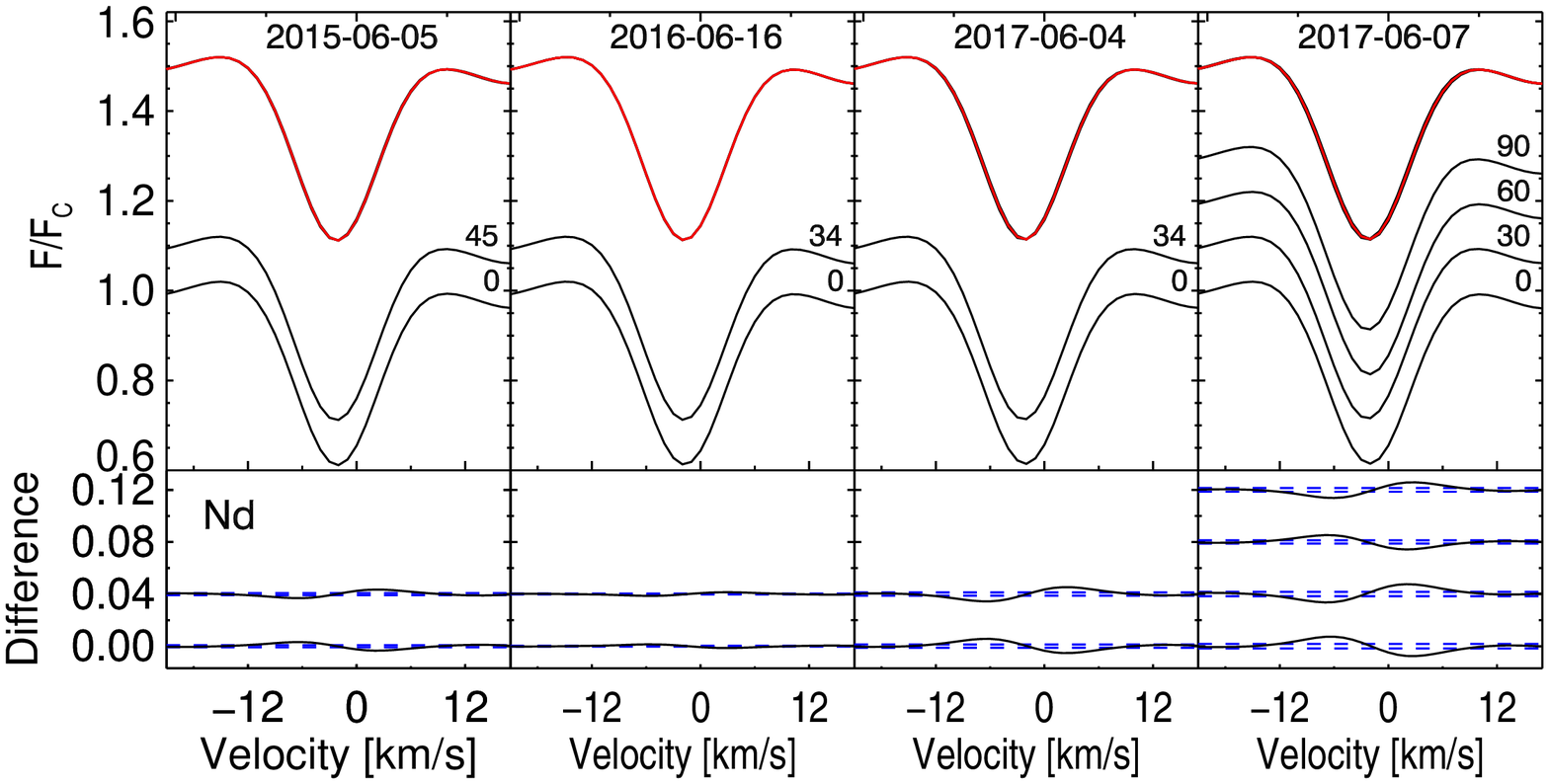}
\includegraphics[width=0.49\textwidth]{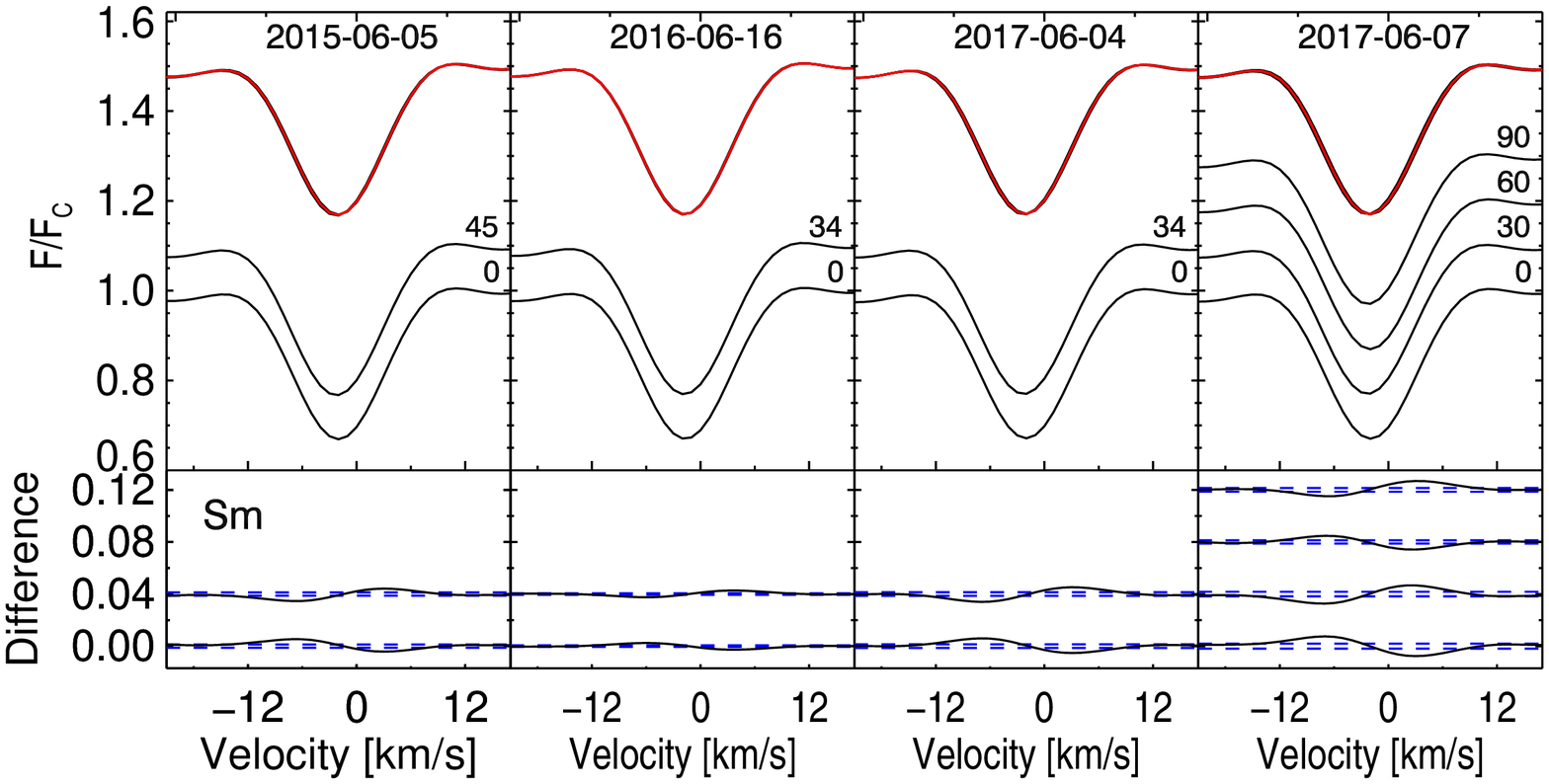}
\caption{
  The LSD line profiles calculated for the line masks All, Fe,
  REE, and several individual REE (La, Ce, Pr, Nd, and Sm).
  \emph{Top panels:} Comparison of the LSD Stokes~$I$ profiles computed for the
  individual subexposures. The individual profiles are shifted vertically for
  better visibility. The upper row shows overplotted profiles together with
  the average profile (red in the online version). The time difference
  in minutes between the individual and the first subexposures is given next
  to the line profiles.
  \emph{Bottom panels:} Differences between Stokes~$I$ profiles computed for
  the individual subexposures and the average Stokes~$I$ profile. The dashed
  lines indicate the standard deviation limits.
         }
   \label{fig:iron}
\end{figure*}

\begin{table}
\centering
\caption{
  Variability of the LSD Stokes~$I$ spectral profiles calculated for each line mask for the sequences of HARPS
  sub-exposures obtained on time scales in the range of 30--90\,min.
  For each line list, the four rows correspond from top to bottom to the observing dates 
  2015 June~5,
  2016 June~16,
  2017 June~4, and
  2017 June~7, respectively. The
  line profile variability in the line cores and in the line wings in \% of
  the normalized flux is presented in Columns~2 and 3, whereas the RV shifts
  in \kms{} are listed in Column~4. For the only observation with four subexposures
  obtained in 2017 June~7  we give the variability ranges. 
}
\label{tab:puls}
\begin{tabular}{lcccc}
\hline
\multicolumn{1}{c}{Line} &
\multicolumn{1}{c}{Core} &
\multicolumn{1}{c}{Wing} &
\multicolumn{1}{c}{RV} \\
\multicolumn{1}{c}{mask} &
\multicolumn{1}{c}{Variability} &
\multicolumn{1}{c}{Variability} &
\multicolumn{1}{c}{} \\
\multicolumn{1}{c}{} &
\multicolumn{1}{c}{(\%)} &
\multicolumn{1}{c}{(\%)} &
\multicolumn{1}{c}{(\kms)} \\
\hline
All       & 0.191          & 0.616 & $-$0.15 \\
          & 0.201          & 0.281 & $-$0.07 \\
          & 0.190          & 0.879 & $-$0.21 \\
          & 0.177\,--\,0.200   & 0.161\,--\,1.150 & $-$0.03\,--\,0.27 \\
\hline
Fe        & 0.157          & 0.699 & $-$0.12 & \\
          & 0.171          & 0.197 & $-$0.04  &\\
          & 0.004          & 0.357 & $-$0.21 & \\
          & 0.049\,--\,0.085   & 0.250\,--\,0.405 & $-$0.08\,--\,$-$0.28 & \\
\hline
REE       & 0.192          & 0.684 & $-$0.14 & \\
          & 0.204          & 0.306 & $-$0.07 & \\
          & 0.200          & 0.919 & $-$0.21 & \\
          & 0.181\,--\,0.203   & 0.171\,--\,1.238 & $-$0.03\,--\,$-$0.28 & \\
\hline
La        & 0.172          & 1.435 & $-$0.09 & \\
          & 0.239          & 0.599 & $-$0.07 & \\
          & 0.135          & 1.093 & $-$0.20 & \\
          & 0.120\,--\,0.197   & 0.268\,--\,1.334 & $-$0.02\,--\,$-$0.24 & \\
\hline
Ce        & 0.397          & 0.471 & $-$0.22 & \\
          & 0.440          & 0.261 & $-$0.10 & \\
          & 0.468          & 0.778 & $-$0.20 & \\
          & 0.373\,--\,0.395   & 0.183\,--\,0.869 & $-$0.03\,--\,$-$0.24  &\\
\hline
Pr        & 0.126          & 0.260 & 0.02 & \\
          & 0.061          & 0.121 & 0.03 & \\
          & 0.104          & 0.394 & $-$0.21 & \\
          & 0.015\,--\,0.096   & 0.066\,--\,0.541 & $-$0.30\,--\,0.02  &\\
\hline
Nd        & 0.309          & 0.689 & $-$0.13  &\\
          & 0.292          & 0.282 & $-$0.06  &\\
          & 0.297          & 1.120 & $-$0.20  &\\
          & 0.245\,--\,0.298   & 0.227\,--\,1.514 & $-$0.03\,--\,$-$0.27  &\\
\hline
Sm        & 0.087          & 1.081 & $-$0.23  &\\
          & 0.056          & 0.546 & $-$0.13  &\\
          & 0.145          & 1.201 & $-$0.27  &\\
          & 0.060\,--\,0.101   & 0.282\,--\,1.484 & $-$0.04\,--\,$-$0.36  &\\
\hline
\end{tabular}
\end{table}

HARPSpol observations are usually split
into either two or four subexposures and allow us to study any changes in the
line profile shape or RV shifts on the time scales corresponding
to the duration of the individual subexposures, which is of the order of 30--45\,min.
We note, however, that since this duration is by a factor of
2.5--3.7 longer than the pulsation period of HD\,101065, we expect that any pulsational variability is smeared over 
the length of our subexposures. We should also keep in mind that the accuracy of the continuum normalization of our 
spectra is limited by photon noise, which is larger for the subexposures compared with the signal-to-noise ratios
of the final combined spectra presented in Table~\ref{tab:log}.
Therefore, in the following, we consider any changes detected at a level of 1\% or less as due 
to instrumental or reduction effects. 
A comparison of the LSD Stokes~$I$ profiles computed for the individual
subexposures obtained on each observing night from 2015 to 2017
for all eight line masks
is displayed in Fig.~\ref{fig:iron}. The presence of very small changes in the LSD Stokes~$I$ line profiles is discovered in 
plots constructed for the line masks All, and rare earth elements.
In Table~\ref{tab:puls} we present our measurements of the LSD line profile variability for each 
observing epoch and all line masks.
The largest variability in the line cores is detected in the cerium lines, up to 0.47\% of the 
normalized flux, while the strongest variability in the line wings up to 1.48\% 
and the largest RV shift up to 0.36\,\kms{} are measured for samarium lines.

\begin{figure}
\centering
\includegraphics[width=0.49\textwidth]{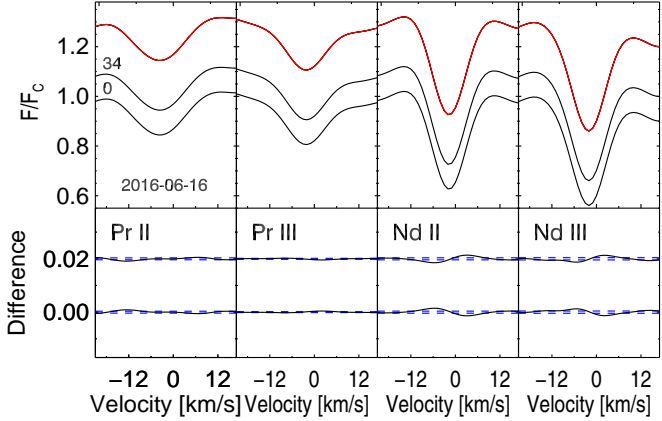}
\caption{
As Fig.~\ref{fig:iron} but using line lists for \ion{Pr}{ii}, \ion{Pr}{iii}, \ion{Nd}{ii}, 
and \ion{Nd}{iii} compiled for observations obtained on 2016 June~16.
         }
   \label{fig:PrNd23}
\end{figure}

As discussed in Sect.~\ref{sec:vert}, the line formation depth of the Pr and  Nd lines in 
the second ionization stage is different compared with that of the Pr and Nd lines in the first ionization stage, i.e.\ 
the \ion{Pr}{iii} and \ion{Nd}{iii} lines are expected to be formed in the upper atmospheric layers.
In Fig.~\ref{fig:PrNd23} we present LSD profiles calculated separately for 
Pr and Nd in different ionization stages. No obvious difference
between the different Pr and Nd ions is indicated in our data, probably due to the long exposure 
times of 34\,min for our subexposures, which exceeds the pulsation period by a factor of 2.8. The variability
of the line profiles belonging to  \ion{Pr}{ii} and \ion{Pr}{iii} is only of the order of 0.1\% of the normalized
continuum.

\begin{figure}
\centering
\includegraphics[width=0.30\textwidth]{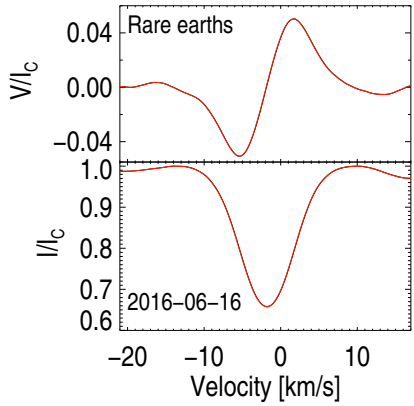}
\caption{
  LSD Stokes~$V$ and $I$ profiles calculated for the REE line list.
  The profiles calculated without taking into account pulsational RV shifts between two subexposures
  are presented by the red color (in the online version) whereas profiles corrected for the RV shift are
  presented in black color.
They are virtually indistinguishable.
           }
   \label{fig:magpuls}
\end{figure}

Pulsations are also known to have an impact on the analysis of the presence
of a magnetic field and its strength by introducing line profile shifts in radial velocity
and shape changes between subexposures
\citep[e.g.][]{Hubrig2011,Jarvinen2017}. Especially in the case
where the duration of the subexposures is comparable to the length of the pulsation period,
the diagnostic null spectra calculated with the
purpose to diagnose spurious polarization, usually do not appear flat.
Since our subexposure duration of the order of 30--45\,min is much
longer than the pulsation period of HD\,101065, we expect that pulsational signatures
in the null spectra are substantially smoothed out. Indeed, as is shown in Fig.~\ref{fig:rot},
only very tiny, hardly recognizable features are present in the null spectra calculated 
for the REE. The amplitude of these features is only of the order of 0.1\% of the
normalized flux. 

To test  the impact of the pulsational variability on our analysis of the magnetic field strength,
the magnetic field measurements were carried out using the left-hand-polarized and right-hand-polarized spectra 
for each subexposure separately, as was already done in previous studies of pulsating 
stars by e.g.\ \citet{Hubrig2011} and \citet{Jarvinen2017}.  The corresponding LSD Stokes~$V$ and $I$ profiles 
calculated for the REE line list using the observations obtained in 2016 are presented in 
Fig.~\ref{fig:magpuls}.   
No significant change in the magnetic field strength was detected in this test, suggesting that our HARPSpol magnetic field 
measurements are not affected by the presence of pulsations. The measurements using the line profiles 
corrected for the RV pulsational shift show an increase of 
the strength of the longitudinal magnetic field by just 21\,G,
which is of the order of the measurement uncertainties of 19\,G.
To summarize, no impact of pulsational variability is detected in our HARPSpol data.

\section{Discussion}
\label{sec:disc}

Our new spectropolarimetric observations of the magnetic roAp star HD\,101065 with HARPSpol over the last three 
years do not indicate any field strength variability within the measurement uncertainties.
Using our measurements along with all available longitudinal magnetic field measurements presented in the literature
and adopting a dipolar structure of the magnetic field, we estimate the probable length of the rotation 
period $P_{rot}\approx188$\,yr. 
This result is certainly subject to the poor sampling of this period.
Our estimation of 
a 188\,yr period should thus not be considered as a unique solution,
given the time coverage of the longitudinal magnetic field measurements of 43\,yr,
which represents a lower limit on the true rotation period of HD\,101065.

Longitudinal magnetic field  measurements using individual masks with lines corresponding to different elements 
reveal distinct differences in the field strengths, which can most likely be ascribed to inhomogeneous 
distributions of these elements in both the horizontal and vertical directions.
The lowest longitudinal magnetic field value was obtained for cerium
($\left<B_{\rm z}\right> \approx -620$\,G)
and the strongest longitudinal magnetic field value ($\left<B_{\rm z}\right> \approx -860$\,G) for neodymium, indicating 
that neodymium lines may form in spots that
are close to the magnetic pole while cerium lines form at some distance from the magnetic pole.
Among the studied elements, Nd and Ce lines are the strongest
in the spectrum of HD\,101065, whereas  Fe and Pr lines appear rather weak.

Longitudinal magnetic field measurements using Nd lines of different ionization stages
reveal the presence of a rather strong magnetic field gradient of the order of 300\,G. Using the line mask 
for the \ion{Nd}{ii} lines we measure a longitudinal magnetic field $\left<B_{\rm z}\right>=-969\pm17$\,G,
whereas using the line mask for the \ion{Nd}{iii} lines we obtain $\left<B_{\rm z}\right>=-663\pm18$\,G. The effect of the
magnetic field gradient is less noticeable in the measurements using the Pr lines. 
Since HD\,101065's atmosphere has an extremely abnormal chemical composition, similar to
a few other roAp stars (e.g. \citealt{Shulyak2010} and references therein), we expect the presence of neodymium  
stratification,  with  the \ion{Nd}{iii} lines formed in the upper atmospheric layer. Thus, the stronger magnetic field
measured using the \ion{Nd}{ii} lines corresponds to a larger atmospheric depth as expected for a dipole
configuration of the magnetic field

Spectropolarimetric monitoring of HD\,101065 in the next tens of years is also important 
for another reason.
In view of the recent work of \citet{Mathys2017}, who
demonstrated that a number of Ap stars with long periods exhibit anharmonicities in their magnetic phase curves,
long rotation periods should be regarded with
caution, as they are the result of an extrapolation based on the assumption that the longitudinal  magnetic
field  variation  curve  does  not significantly depart from a sinusoid. 
Apart from monitoring the strength of the longitudinal magnetic field of HD\,101065
using circularly polarised spectra, a useful approach to
establish the presence of an extremely slow rotation could be to measure broad-band
linear polarisation, which is caused by different saturation of the $\pi$ and
$\sigma$ components of a spectral line in the presence of a magnetic field.
This differential effect is qualitatively similar for all lines, so that in
broad-band observations the contributions of all lines add up. A model for
the interpretation of such observations was developed by \citet{Landolfi1993}.

According to \citet{Mathys2015}, there exist strong indications that the longest periods of Ap stars must
reach about 300\,yr, and even longer, up to 1000\,yr. A rotation period of about 35462.5\,d ($\sim$97\,yr) 
was suggested by \citet{Bychkov2016} for the Ap star $\gamma$~Equ (=HD\,201601). Also the observations 
of broad-band linear polarization by \citet{Leroy1994} over about three years indicated that the 
rotation period of $\gamma$~Equ is longer than 70\,yr.
Six other Ap stars were suggested to possess a rotation period longer than 10\,yr. \citet{Metlova2014} showed that a
rotation period of 7961.8\,d ($\sim$22\,yr) is consistent with 
all available photometric and spectropolarimetric observations of GY\,And (=HD\,9996).  
Magnetic field monitoring of HD\,965 by \citet{Romanyuk2015} indicated a rotation period of about 13\,yr,
while the inclusion of all available longitudinal magnetic field values in the period search suggested a 
period of about 17\,yr (Bychkov et al. 2018, in preparation).
As reported by \citet{Hubrig2002}, HD\,965 discloses a very similar spectral appearance to HD\,101065, showing
a very complex spectrum rich with lines of lanthanides.
\citet{Mathys2017} combined for 33\,Lib (=HD137949) all available longitudinal magnetic field measurements 
from the work of \citet{Mathys1997} and \citet{Romanyuk2014}, reporting that a peak standing
out rather clearly in the periodogram corresponds to a period of 5195\,d ($\sim$14\,yr).
Two more Ap stars, HD\,110066 and HD\,41403, were reported to have long rotation periods,
but the spectropolarimetric observations have very incomplete phase coverage. Using spectrophotometric 
observations, \citet{Adelman1981} suggested for HD\,110066 a period of about 13.5\,yr. The rotation period 
of HD\,41403 of $\sim13$\,yr (Bychkov et al. 2018, in preparation) 
is based on the work of \citet{Kudryavtsev2006}.  
A detailed discussion of the longitudinal magnetic field measurements and the phase curves for the stars 
HD\,201601, HD\,965, HD\,9996, HD\,137949, and HD\,110066 is presented in the work of \citet{Mathys2017}.
This work also confirms the previous result of the study of \citet{wolff1975} that the properties
of stars (chemical composition, magnetic field strength, etc.)  with long periods are similar to those of 
Ap stars with shorter periods.

The processes playing a role in achieving the slow rotation of magnetic Ap stars are not identified yet. It is 
generally assumed 
that Ap stars are slow rotators because of magnetic braking and that most of the angular momentum is lost in the  
pre-main-sequence phase. Using accurate Hipparcos parallaxes for magnetic Ap stars with masses below 3\,$M_\odot$, 
\citet{Hubrig2000a,Hubrig2007}  
showed that  magnetic stars are concentrated towards the centre of the main-sequence band, whereas normal A stars 
occupy the whole width of the main sequence, without a gap. According to these results, there must exist progenitors of
Ap stars in the form of normal A stars, which probably rotate slowly. It is conceivable that the Ap progenitors 
contain a strong field in their interior, which only appears at the surface at some evolutionary stage (e.g.\ \citealt{Nordlund2006}).
A study of projected
rotational velocities in young and old A stars involving a sample of 160 A-type stars with very accurate Hipparcos 
parallaxes did not reveal any statistically significant difference 
between the $v\,\sin\,i$ distributions \citep{Hubrig2000b}. The obtained results 
were also fully compatible with conservation of angular momentum.
It is quite possible that the sample of 160 A-type stars was not representative enough for such a study and
that most of the angular momentum is indeed lost in the pre-main-sequence phase, where the magnetic
field responsible for braking the pre-main-sequence star's rotation is buried under the surface of the star - allowing it to rotate
slowly while staying chemically normal - and appears at the
surface again after the star has completed a part of its life on the main sequence.

The magnetic pre-main-sequence Herbig Ae/Be stars are frequently considered as progenitors of magnetic Ap/Bp stars as
their incidence
of $\sim$7\% is similar to that in magnetic Ap/Bp stars \citep{Alecian2017}. Magnetic fields in Herbig stars might be 
fossils of the early star formation epoch, in which the magnetic field of the parental magnetized core was compressed 
into the innermost regions of the accretion disks (e.g.\ \citealt{Banerjee2006}). 
On the other hand,  \citet{Cauley2014}
studied the \ion{He}{i} $\lambda$10\,830 morphology in a sample of 
56~Herbig~Ae/Be stars and suggested that Herbig~Be stars do not accrete material from their inner
disks in the same manner as classical T\,Tauri stars, which accrete material via magnetospheric accretion.
Only late type Herbig~Be and Herbig~Ae stars show evidence for magnetospheric accretion. In magnetospheric accretion
models the disk material is channeled from the disk's inner edge onto the star along the magnetic field lines implying that 
magnetic fields do exist on the surface of late type Herbig~Be and Herbig~Ae stars. Therefore, the low incidence
of magnetic fields in these stars reported by \citet{Alecian2017} is difficult to understand, but can probably be explained
by the weakness of their magnetic fields \citep{Hubrig2015}.
Obviously, explanations for the magnetic field origin and magnetic field evolution from the PMS stage to the end
of the main-sequence life can be made only through  careful  consolidation  of  magneto-hydrodynamic
theory and observations. 

In conclusion, the presence of a very long rotation period in HD\,101065 along with the previous
detection of slow rotation in six other magnetic Ap stars stars indicates that there possibly exist
other not yet identified slowly rotating Ap stars.
Clearly, the detection and  monitoring of such stars and the characterization of their stellar properties by future observations
is extremely important for our understanding of their formation and the processes playing a role in achieving such 
super-slow rotation. 

\section*{Acknowledgments}
The authors thank the anonymous referee for useful comments
and C.~R.~Cowley for a valuable discussion.
Based on observations made with ESO Telescopes at the La Silla Paranal
Observatory under programme IDs
69.D-0210(A),
270.D-5023(A),
079.D-0240(A), 
191.D-0255(I),
097.C-0277(A), and
099.C-0081(A).
This work has made use of the VALD database, operated at Uppsala
University, the Institute of Astronomy RAS in Moscow, and the University of
Vienna.

\bsp	
\label{lastpage}
\end{document}